\documentclass{doublecol-new}
\usepackage[authoryear]{natbib}
\usepackage{stfloats}
\usepackage{mathrsfs}
\usepackage{graphicx}
\usepackage{subfigure}
\usepackage{amsmath}
\usepackage{bm}
\usepackage{comment}

\theoremstyle{TH}{

}

\theoremstyle{THrm}{

}

\theoremstyle{THhit}{

}

\makeatletter
\def\theequation{\arabic{equation}}

\makeatother

\begin{document}%

\setcounter{page}{1}


\RRH{Six-DoF Stewart Platform Motion Simulator Control using Switchable Model Predictive Control}

\VOL{x}

\ISSUE{x}

\PUBYEAR{xxxx}

\BottomCatch


\PUBYEAR{201X}

\subtitle{}

\title{6DoF Stewart Motion Platform Control using Switchable Model Predictive Control}

%
\authorA{Jiangwei Zhao}
\affA{School of Information Engineering, Pingdingshan University, China\\
E-mail: 6507@pdsu.edu.cn}

\authorB{Zhengjia Xu}
\affB{School of Aerospace, Transport and Manufacturing, Cranfield University, UK \\
E-mail: billy.xu@cranfield.ac.uk}

\authorC{Dongsu Wu}
\affC{College of Civil Aviation, Nanjing University of Aeronautics and Astronautics \\
E-mail: tissle@nuaa.edu.cn }

\authorD{Yingrui Cao}
\affD{School of Information Engineering, Pingdingshan University, China\\
E-mail: 1091939954@qq.com}

\authorE{Jinpeng Xie}
\affE{School of Information Engineering, Pingdingshan University, China\\
E-mail: maomao200703@163.com}

\begin{abstract}

Due to excellent mechanism characteristics of high rigidity, maneuverability and strength-to-weight ratio, 6 Degree-of-Freedom (DoF) Stewart structure is widely adopted to construct flight simulator platforms for replicating motion feelings during training pilots. Unlike conventional serial link manipulator based mechanisms, Upset Prevention and Recovery Training (UPRT) in complex flight status is often accompanied by large speed and violent rate of change in angular velocity of the simulator. However, Classical Washout Filter (CWF)  based Motion Cueing Algorithm (MCA) shows limitations in providing rapid response to drive motors to satisfy high accuracy performance requirements. This paper aims at exploiting Model Predictive Control (MPC) based MCA which is proved to be efficient in Hexapod-based motion simulators through controlling over limited linear workspace. With respect to uncertainties and control solution errors from the extraction of Terminal Constraints (COTC), this paper proposes a Switchable Model Predictive Control (S-MPC) based MCA under model adaptive architecture to mitigate the solution uncertainties and inaccuracies. It is verified that high accurate tracking is achievable using the MPC-based MCA with COTC within the simulator operating envelope. The proposed method provides optimal tracking solutions by switching to MPC based MCA without COTC outside the operating envelope. By demonstrating the UPRT with horizontal stall conditions following Average Absolute Scale(AAS) evaluation criteria, the proposed S-MPC based MCA outperforms MPC based MCA and SWF based MCA by 42.34\% and 65.30\%, respectively.
\end{abstract}

\KEYWORD{Model Adaptive Architecture; Motion Cueing Algorithm; Model Predictive Control; Steward Motion Platform Control.}

\maketitle

\section{Introduction}
\begin{figure*}[htbp]
\centering
\caption{Stewart flight simulator motion platform.}
\includegraphics[scale=0.5]{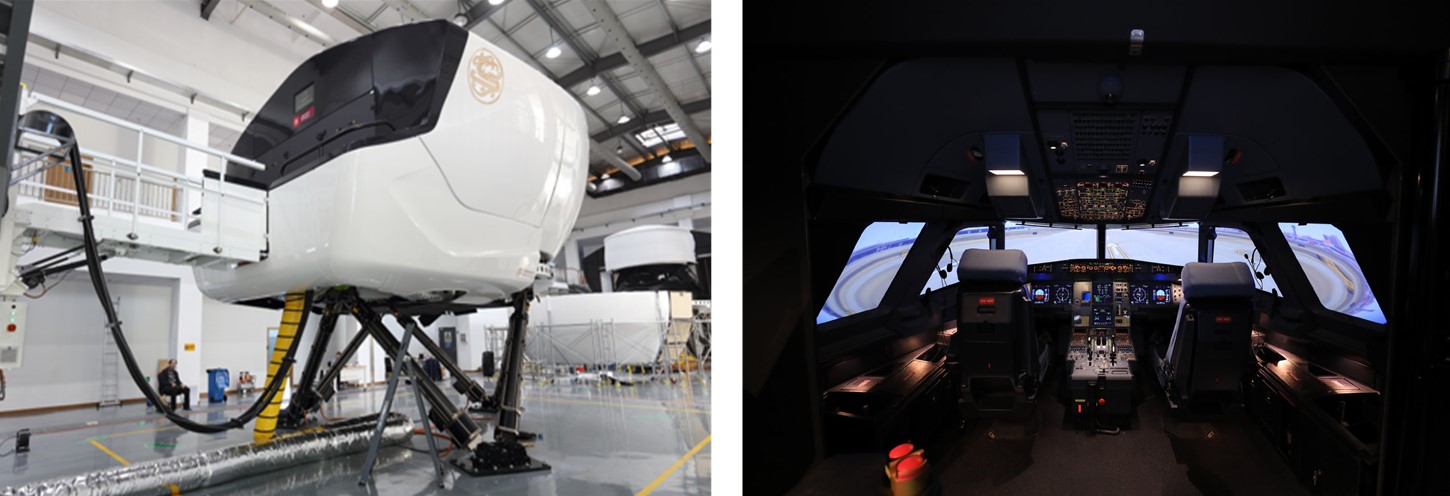}
\label{fig:steward}
\end{figure*}

A flight simulator is a sophisticated device engineered to replicate the flight conditions of an aircraft, providing an authentic emulation of the aircraft's behaviour and interaction with its environment. Owing to its superior mechanical properties, including high rigidity, exceptional manoeuvrability, and an optimal strength-to-weight ratio, the 6 Degrees of Freedom (DoF) Stewart platform is extensively employed in the construction of flight simulator platforms (as illustrated in Figure \ref{fig:steward}).

To simulate aircraft motion using Stewart platform, \cite{schmidt1970motion} designed a Classical Washout
Filter (CWF) based Motion Cueing Algorithm (MCA), encompassing translational, rotational, and tilt coordination channels. \cite{asadi2016particle} applied a particle swarm optimisation algorithm for the operation of running the CWF based MCA, making better use of workspace limitations. Adaptive MCA is proposed via dynamically adjusting CWF based MCA parameters to maximise the repeatability towards reference signals like \cite{parrish1975coordinated,parrish1973coordinated}. Furthermore, \cite{baghaee2018performance,muyeen2013modeling,zadeh1988fuzzy,qazani2020adaptive,qazani2021adaptive,asadi2022adaptive} worked on applying fuzzy logic control theory into MCA to formulate adaptive MCA problem where human vestibular system is capable of replicating the true motion feeling \cite{asadi2019increasing,asadi2015incorporating,hwang2009adaptive}. With respect to human vestibular models, Linear Quadratic Regulator (LQR) is a prevalent approach reproducing human feelings through integrating the human vestibular model into the optimal MCA problem formulation \cite{kang2022ldct,qazani2021optimal,asadi2016robust,cleij2020optimizing,qazani2020optimising,qazani2020motion} owing to benefits of leveraging control and system performance \cite{zhuangkun2024}. Nevertheless, it is found that the adjustment only exists in the parameter space domain, lacking the dynamic adjustment features in the systematic parameter domain that may introduce significant errors resulting from modelling uncertainties.

To further enhance the MCA performance in terms of enhancing flexibility in the systematic parameter domain, \cite{asadi2015optimisation,dagdelen2009model,qazani2021time,ploeg2020sensitivity,asadi2023novel,khusro2020mpc} studied on Model Predictive Control(MPC) based MCA which incorporates human vestibular system and flight simulation manoeuvring dynamics system models into the MCA modelling, forming the truly optimal control MCA that achieves optimal control purposes. \cite{cleij2019comparison} paid attention to comparing the MPC based MCA and the CWF based MCA in practice, validating significant performance improvement by incorporating MPC into MCA. Given the leg length constraints in Stewart platforms, \cite{qazani2020motion} presented a linear MPC based MCA with Considering of Terminal Conditions (COTC) for addressing physical constraint problems.

\begin{figure}[htbp]
\centering
\subfigure[MPC based MCA longitudinal acceleration trajectory tracking curve]{
\begin{minipage}[t]{1\linewidth}
\centering 
\label{fig:mpc_g}
\includegraphics[width=2.6in]{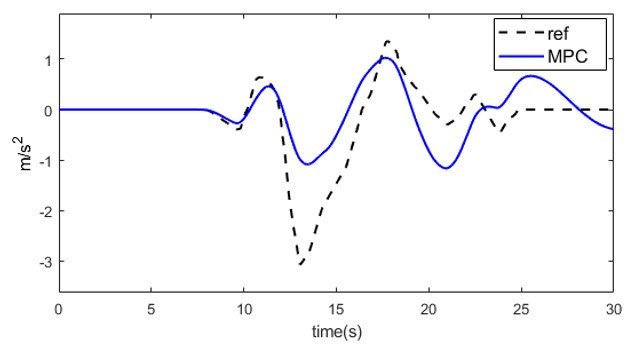}
\end{minipage}%
}%

\subfigure[S-MPC based MCA longitudinal acceleration trajectory tracking curve]{
\begin{minipage}[t]{1\linewidth}
\centering 
\label{fig:smpc_g}
\includegraphics[width=2.6in]{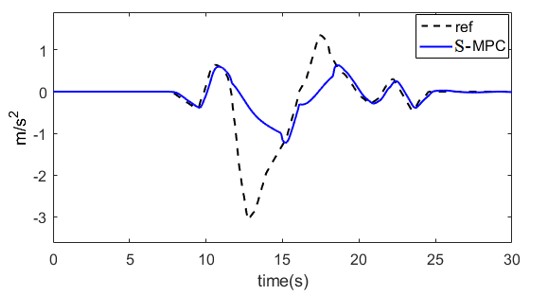}
\end{minipage}%
}%
\caption{Overall MPC and S-MPC based MCA.}
\label{fig:mpc_smpc}
\end{figure}

As studied by multiple studies \cite{asadi2015optimisation,dagdelen2009model,qazani2021time,ploeg2020sensitivity,asadi2023novel,khusro2020mpc}, the implementation of MPC in MCA can lead to destabilisation of prior models under complex operational scenarios. These scenarios include bumping, stalling, and platform fluctuations, primarily because of inadequate terminal conditions concerning weight and state parameters. This deficiency is visually represented in Figure \ref{fig:mpc_g} and is known to substantially degrade the fidelity of motion simulations.

A novel Switchable Model Predictive Control (S-MPC) based MCA is proposed, demonstrating promising outcomes in simulations of complex flight scenarios, as depicted in Figure \ref{fig:smpc_g}. Within the operational boundaries of the simulator, utilising an S-MPC based MCA integrated with a COTC ensures realistic motion simulation outcomes. Conversely, outside these boundaries, an S-MPC-based MCA devoid of COTC is designed to achieve optimal approximation tracking. Furthermore, the development of innovative switchable mechanisms is proposed to mitigate the effects of controller transitions, enhancing the overall system stability. Evaluation of S-MPC performance is performed on the analysis of terminal states \cite{kwon1977modified,kwon1978feedback}, the implementation of an infinite output prediction horizon \cite{kouvaritakis2016model}, and the optimisation of the terminal weighting matrix (TWM) \cite{kwon1983stabilizing,kwon1989receding}.

The rest of the paper falls into the following structures. Section \ref{section2} describes the MPC system predictive model,including the human vestibular motion perception model and kinematic model of aircraft simulator. The S-MPC based MCA with COTC and without COTC are explained in Section \ref{section3}. In Section \ref{section4}, the outcomes show that the S-MPC based MCA has better motion simulation performance and is capable of making better use of the platform's working space versus the CWF based MCA and MPC based MCA. Section \ref{section5} concludes the primary findings.

\section{System Prediction Model}\label{section2}

\subsection{Human Vestibular System Model}
The organ that provides the primary source of information for inertial signals such as gravity is the vestibular system, encompassing semicircular canals and otoliths.

The semicircular canals, the sensory organs primarily responsible for perceiving rotational motion feelings, are composed of liquid-filled semicircular tubes located approximately in the three orthogonal axes of the human head. The semicircular canals can perceive rotational motion signals, which are also known as the angular velocity sensors of the human body. Equation (1) gives the transfer function of the linear semicircular canal model:

\begin{equation}
\frac{\hat{\omega}}{\omega} = \frac{T_L T_a s^2}{(T_L s + 1)(T_a s + 1)(T_S s + 1)}
\end{equation}
Where $\omega$ is the input angular velocity signal in the vestibular model. $\hat{\omega}$ stands for the human-perceived angular velocity. $T_L$ and $T_a$ represent the model parameters of the semicircular canals.

Similar to the semicircular canal, the otolith organ is modelled to perceive  translational motion feelings, and its transfer function is defined by equation (2):

\begin{equation}
\frac{\hat{a}}{a} = \frac{\Gamma_a s + 1}{\Gamma_L s + 1} \cdot \frac{K}{\Gamma_s s + 1}
\end{equation}
Where $a$ is the input translational acceleration signal of the vestibular model, $\hat{a}$ is the translational acceleration perceived by the human body. $\Gamma_a$, $\Gamma_L$ and $K$ are the model parameters of the otolith.

\begin{figure}[htbp]
\centering
\caption{Schematic diagram of coordination transfer between horizontal and vertical tilt of simulator.}
\includegraphics[scale=0.65]{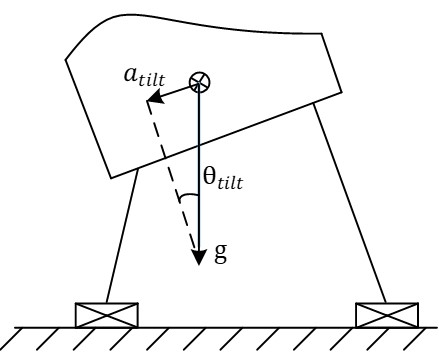}
\label{fig:Schematic}
\end{figure}

Providing limited motion workspace in simulator platforms, its sustained acceleration motion is calculated by the tilt coordination effect, with the aid of the lateral and longitudinal components of the gravitational acceleration \cite{grant1984governing}. The goal of tilt coordination is hereby achieved by converting the low-frequency component of the acceleration signal into the tilt angle of the simulator platform as illustrated in Fig. \ref{fig:Schematic}.

Therefore, the formula of calculating the translational acceleration of the tilt coordination with respect to the tilt angle is derived below:

\begin{equation}
\theta_{\text{tilt}} = \arcsin\left(\frac{a_{\text{tilt}}}{g}\right) \approx \frac{a_{\text{tilt}}}{g}
\end{equation}
Where $\theta_{\text{tilt}}$ represents the angle of tilt coordination, $a_{\text{tilt}}$ denotes the translational acceleration component resulting from tilt coordination, and $g$ denotes the gravitational acceleration.

The transfer function for tilt coordination is defined as follows:

\begin{equation}
\frac{\hat{a}_t}{\omega_{\text{tilt}}} = \frac{g \cdot K \left(\Gamma_a s + 1\right)}{s \left(\Gamma_L s + 1\right) \left(\Gamma_s s + 1\right)}
\end{equation}
where $\omega_{tilt}$ represents the input angular velocity signal from the platform tilt of the vestibular model, and $\hat{a}_t$ represents the translational acceleration felt by the human body.

Modelling was conducted on the complete vestibular system through integrating the state-space models pertaining to the semicircular canal, otolith organ and tilt coordination, resulting in the following system:

\begin{equation}
x_P = \mathbf{A}_p x_p + \mathbf{B}_p u_p
\end{equation}
\begin{equation}
y_p = \mathbf{C}_p x_p
\end{equation}
where $x_p=\left[x_{oth}\ x_{tilt}\ x_{acc}\right]^T\in\mathbb{R}^{21\times1}$ denotes the state vector in the human vestibular model, $x_{oth}$, $x_{tilt}$, and $x_{acc}$ are the state vector of otolith organ, tilt coordination, and semicircular canal model, respectively. $u_p=\left[a_p\ \ \omega_{P,tilt}\ \ \omega_{P,rot}\ \right]^T\in\mathbb{R}^{8\times1}$represents the input vector. $a_p$, $\omega_{P,tilt}$, and $\omega_{P,rot}$ are the vector of platform translational acceleration, tilt coordination angular velocity, and angular velocity, respectively. $y_p=\left[\ {\hat{a}}_P\ \ {\hat{a}}_{P,tilt}\ \ {\hat{\omega}}_{P,rot}\right]^T\in\mathbb{R}^{8\times1}$ is the output vector. ${\hat{a}}_P$, ${\hat{a}}_{P,tilt}$, and  ${\hat{\omega}}_{P,rot}$ are the output vector of platform translational acceleration, translational acceleration generates by tilt coordination, and angular velocity, respectively. $\mathbf{A}_p\in\mathbb{R}^{21\times21}$, $\mathbf{B}_p\in\mathbb{R}^{21\times8}$, and $\mathbf{C}_p\in\mathbb{R}^{8\times21}$ are integration state-space model matrices.

\subsection{Motion Platform Model}
\begin{figure}[htbp]
\centering
\caption{Geometric structure and reference coordinate system of simulator platform.}
\includegraphics[scale=0.18]{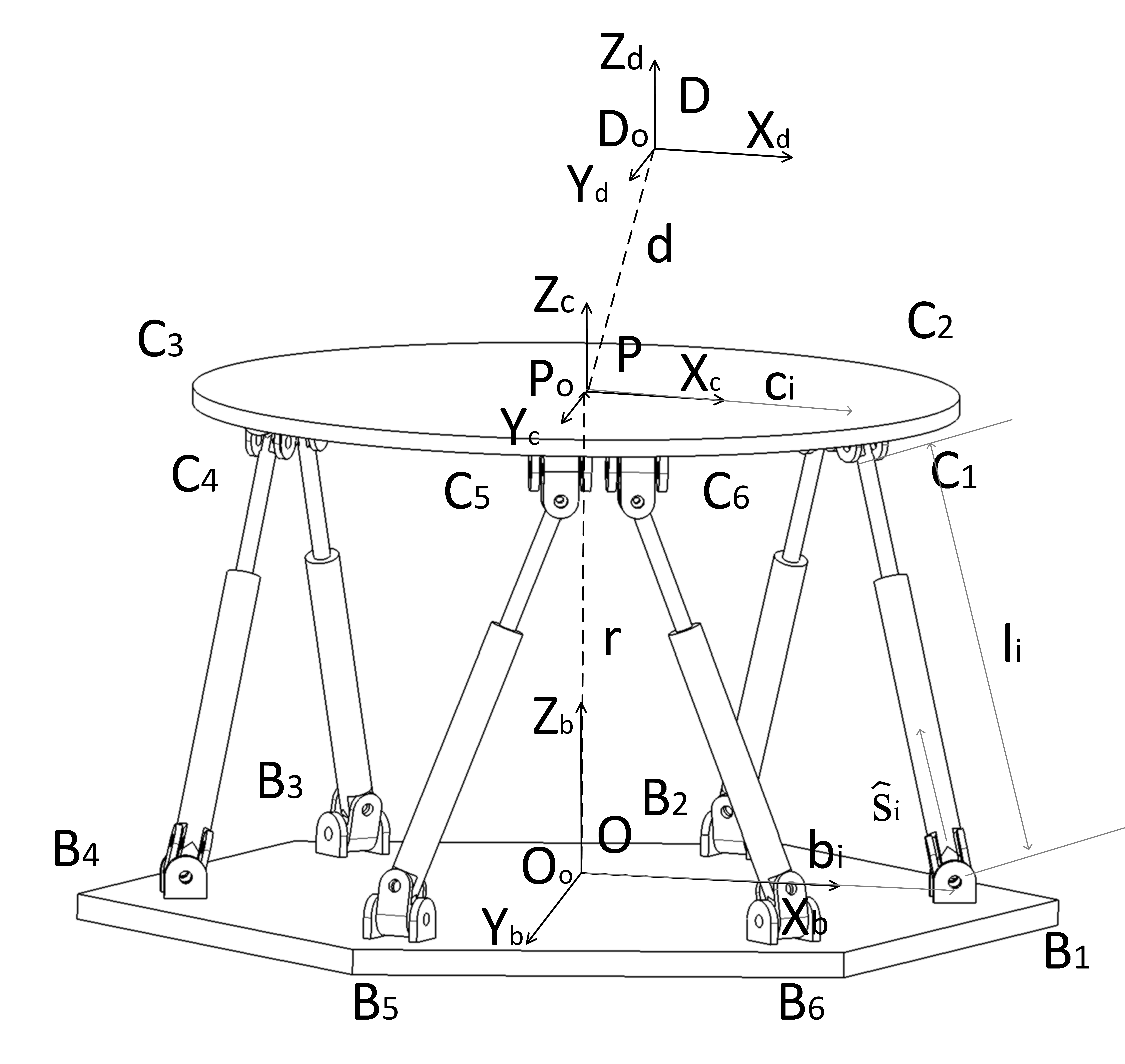}
\label{fig:geo}
\end{figure}

\begin{figure*}[htbp]
\centering
\caption{S-MPC integrated MCA system diagram.}
\includegraphics[scale=0.7]{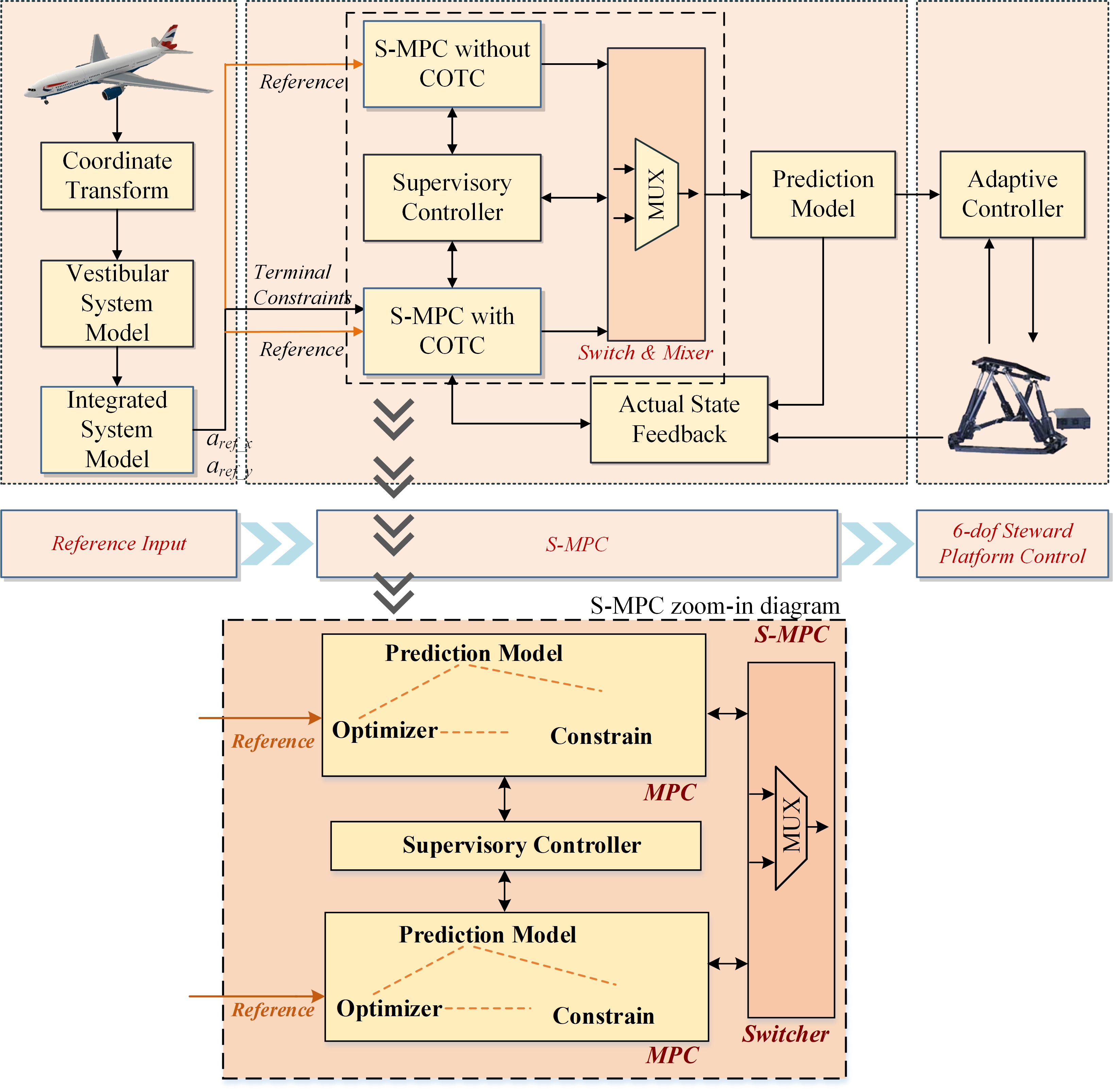}
\label{fig:smpc}
\end{figure*}

The geometric structure and reference inertial frame of the Stewart platform are illustrated in Fig. \ref{fig:smpc}, presenting the coincidence between the coordinate origin and the platform’s fixed base centroid (Point $O_0$ in Fig. \ref{fig:smpc}), which does not follow the motion platform movement. The platform frame is fixed on the platform’s centre of mass (Point $P_0$ in Fig. \ref{fig:geo}) and moves together with the motion platform. The operator frame is fixed to the operator’s head, making the corresponding movements. The coordinate origin coincides with the operator’s eye point, and the corresponding coordinate system is represented by $D_0$.

The inverse kinematic relations with regard to the motion platform are taken into account to constrain the velocity and length of actuator, which combine the actuator kinematic relations proposed by \cite{harib2003kinematic}. With the aim of obtaining the leg length, a closed loop equation for each mounted point $\bm{c}_i$ towards individual leg $l_i$ is derived as:

\begin{equation}
    \bm{c}_i = \bm{r} + \bm{R}_P^O \cdot \bm{c}_i^P
\end{equation}
where $\bm{r}$ represents the position vector of $O_P$ on the motion platform in the inertial frame. $\bm{c}_i$ and $\bm{c}_i^P$ are the position vector of the connection point between the platform and the legs in the inertial frame, and that between the platform and the legs in the platform coordinate frame, respectively.

$\bm{R}_O^P$ denotes the rotation matrix from the fixed motion platform coordinate to the original motion platform coordinate:

\begin{equation*}
    \bm{R}_P^O = \begin{bmatrix}
\boldsymbol{c}\psi \boldsymbol{c}\phi - \boldsymbol{c}\theta \boldsymbol{s}\phi \boldsymbol{s}\psi & -\boldsymbol{s}\psi \boldsymbol{c}\phi - \boldsymbol{c}\theta \boldsymbol{s}\phi \boldsymbol{c}\psi & \boldsymbol{s}\theta \boldsymbol{s}\phi \\
\boldsymbol{c}\psi \boldsymbol{s}\phi + \boldsymbol{c}\theta \boldsymbol{c}\phi \boldsymbol{s}\psi & -\boldsymbol{s}\psi \boldsymbol{s}\phi + \boldsymbol{c}\theta \boldsymbol{c}\phi \boldsymbol{c}\psi & -\boldsymbol{s}\theta \boldsymbol{c}\phi \\
\boldsymbol{s}\psi \boldsymbol{s}\theta & \boldsymbol{c}\psi \boldsymbol{s}\theta & \boldsymbol{c}\theta
\end{bmatrix}
\end{equation*}
where $\boldsymbol{c}$ and $\boldsymbol{s}$ denote cosine and sine functions. $\phi$, $\theta$ and $\psi$ respectively denote the roll, pitch and yaw angle.

The closed loop equation for each leg vector $l_i$ is reformulated as:

\begin{equation}
    \bm{L}_i = l_i \cdot \hat{\bm{s}}_i = \bm{c}_i - \bm{b}_i = \bm{r} + \bm{R}_P^O \cdot \bm{c}_i^P - \bm{b}_i
\end{equation}
where $\bm{L}_i$ represents the vector of the i-th leg. $l_i$ represents the length of the i-th leg. $\hat{\bm{s}}_i$ represents the unit direction vector of the i-th leg. $\bm{b}_i$ is the position vector of the connection point between the base and the leg in the inertial frame.

Furthermore, the calculation of the rate of change in leg length can be derived from the dot product of the leg function in the closed loop format from equation (9), and the expression after taking the derivative is calculated as follows:
\begin{equation}
\dot{l}_i = \frac{d}{dt} \left( \sqrt{\bm{L}_i \cdot \bm{L}_i} \right)
\end{equation}

$\dot{l}_i$ is a nonlinear function formed by two factors of the platform linear velocity $\bm{v}_p$ and angular velocity $\bm{\omega}_p$, hereby $\dot{l}_i$ is expressed as:

\begin{equation}  
\dot{l}_i = h(\bm{x}_l)
\end{equation}  
where $\bm{x}_l = \begin{bmatrix} \bm{v}_p & \bm{\omega}_p \end{bmatrix}^T \in \mathbb{R}^{6 \times 1}$ represents the motion and velocity vector of the platform.

The nonlinear equation (11) is further linearised at the original coordinate $O_0$:
\begin{equation}  
\dot{l}_i = \bm{A}_i \bm{v}_p + \bm{B}_i \bm{\omega}_p
\end{equation}  

\subsection{Integrated System Model}
The integrated system model includes a vestibular system model with tilted coordination and a kinematic model designed for the simulator motion platform. The state equation of the integrated system in the inertial frame of the simulator is shown as follows:

\begin{equation}
\dot{\mathbf{x}}_t = \begin{cases}
\dot{r}_p = v_p \\
\dot{v}_p = a_p \\
\dot{\beta}_{p,\text{rot}} = \omega_{p,\text{rot}} \\
\dot{\beta}_{p,\text{tilt}} = \omega_{p,\text{tilt}} \\
\dot{x}_p = \mathbf{A}_p x_p + \mathbf{B}_p \left[ a_P^D ; \omega_{P,\text{tilt}}^D ; \omega_{P,\text{rot}}^D \right] \\
\dot{l}_i = \mathbf{A}_i v_p + \mathbf{B}_i \omega_p
\end{cases}
\end{equation}

\begin{equation}
    a_P^D = a_p + \bm{v}_p \times \left( \bm{R}_0^P \bm{q}^P \right) + \bm{\omega}_p \times \bm{\omega}_p \times \left( \bm{R}_0^P \bm{q}^P \right)
\end{equation}

\begin{equation}
    \omega_{P,\text{rot}}^D = \bm{C}_D^0 \omega_{p,\text{rot}}
\end{equation}

\begin{equation}
    \omega_{P,\text{tilt}}^D = \bm{C}_D^0 \omega_{p,\text{tilt}}
\end{equation}

\begin{equation}
    \omega_p = \omega_{p,\text{rot}} + \omega_{p,\text{tilt}}
\end{equation}
In which, $\bm{r}_p$, $\bm{v}_p$, $\bm{a}_p$, $\bm{\beta}_p$, and $\bm{\omega}_p$ represent the displacement vector, translational velocity vector, translational acceleration vector, angular displacement vector, and angular velocity vector of the simulator motion platform in the platform frame, respectively. $\bm{a}_P^D$, $\bm{\omega}_{P,\text{tilt}}^D$, and $\bm{\omega}_{P,\text{rot}}^D$ represent the translational acceleration vector, tilt coordination angular velocity, and angular velocity vector at the operator eyepoint in the operator frame, respectively.

The equation below explains the discretized MPC state-space model:
\begin{equation}  
\bm{x}_m(k + 1) = \bm{A}_m \bm{x}_m(k) + \bm{B}_m \bm{u}(k) 
\end{equation}  
\begin{equation}  
\bm{y}(k) = \bm{C}_m \bm{x}_m(k) 
\end{equation}  
$\bm{x}_m = \left[ \bm{r}_p \ \bm{v}_p \ \bm{a}_p \ \bm{\beta}_{p,\text{rot}} \ \bm{\beta}_{p,\text{tilt}} \ \bm{x}_p \ l_i \right]^T$ and $\bm{y}$ denote the state vector and output vector of the MPC model. Also, $\bm{A}_m$, $\bm{B}_m$ and $\bm{C}_m$ are the model's final matrices; and $\bm{u}(k)$ is the input vector, including $\left[ \bm{a}_p \ \bm{\omega}_{p,\text{rot}} \ \bm{\omega}_{p,\text{tilt}} \right]^T$. Let us define the iterative state-space model of (18,19) as:

\begin{equation}  
\bm{x}(k + 1) = \bm{A} \bm{x}(k) + \bm{B} \Delta \bm{u}(k) 
\end{equation}  

\begin{equation}  
\bm{y}(k) = \bm{C} \bm{x}(k) 
\end{equation}  

In above equations, the control input function $\Delta \bm{u}(k) = \bm{u}(k) - \bm{u}(k-1)$, the state function $\bm{x}(k) = \left[ \bm{x}_m(k) - \bm{x}_m(k-1) \ \bm{y}(k) \right]^T$. The matrices $\bm{A}$, $\bm{B}$, and $\bm{C}$ are defined by:

\begin{equation}  
\bm{A} = \begin{bmatrix}
\bm{A}_m & \bm{0} \\
\bm{C}_m \bm{A}_m & \bm{I}
\end{bmatrix}, \quad \bm{B} = \begin{bmatrix}
\bm{B}_m \\
\bm{C}_m \bm{B}_m
\end{bmatrix}, \quad \bm{C} = \begin{bmatrix}
\bm{0} & \bm{I}
\end{bmatrix} 
\end{equation}  

\section{S-MPC based MCA}\label{section3}
The S-MPC based MCA is specifically designed to enhance the dynamic simulation capabilities of aircraft simulators, particularly for complex flight states. It effectively addresses the challenges posed by unexpected fluctuations on the simulator platform, resulting from insufficient terminal conditions in the conventional MPC-based MCAs. The proposed S-MPC based MCA architecture comprises the following key components: $\bullet$ an MPC-based MCA with  COTC, $\bullet$ an MPC-based MCA without COTC, $\bullet$ an Adaptive Weight Regulator (AWR), $\bullet$ a Supervisory Controller (SC), and $\bullet$ a Switch Mixer (SM).

\subsection{S-MPC based MCA System Architecture}

The system architecture of the S-MPC based MCA is depicted in Figure \ref{fig:smpc}. The SC plays a pivotal role in monitoring the status of the two controllers and executing channel switching under predefined conditions. The criteria for transitioning from a model predictive controller with COTC to one without COTC hinges on the availability of a solution from the former. In the absence of a solution, switching to the model predictive controller without COTC becomes necessary. Conversely, when lateral or longitudinal acceleration tracking errors are minimal and within a predefined neighbourhood, the system reverts to the model predictive controller with COTC.

Due to abrupt state changes during transitions, switching between controllers could induce a pronounced sense of movement leading to performance divergence. To smooth the performance during switching, a time-varying weighting approach is employed in SC to facilitate a smooth transition between states, thereby softening the switching impact and enhancing the user experience.

\subsection{Controller design for  MPC based MCA with COTC }
The cost function $J(k)$ is to reduce the motion sensation error and replicate real aircraft pilot feelings. Moreover, the reference signal $\bm{x}_{\text{ref}}$ is defined and integrated into the objective function. 

$\bm{x}(t + n|t)$ indicates the state at the $n$-th time step ($T_s$). The optimisation problem of the MPC-based MCA with COTC is presented:

\begin{multline} 
\min_{\substack{u(i), x(k) \in U}} J(k) = \sum_{i=1}^{N_p-1} \left( \bm{x}_{\text{ref}}(k + i \mid k) - \bm{x}(k + i \mid k) \right)^T 
\\ \bm{Q} \left( \bm{x}_{\text{ref}}(k + i \mid k) - \bm{x}(k + i \mid k) \right) + \sum_{i=1}^{N_c-1} \bm{u}^T(k + i \mid k) \\
\bm{S} \bm{u}(k + i \mid k) + \sum_{i=1}^{N_c-1} \Delta \bm{u}^T(k + i \mid k) \bm{R} \Delta \bm{u}(k + i \mid k) + \\
\left( \bm{x}_{\text{ref}}(k + N_p \mid k) - \bm{x}(k + N_p \mid k) \right)^T \bm{Q}_{N_p} ( \bm{x}_{\text{ref}}(k + N_p \mid k) \\- \bm{x}(k + N_p \mid k) ) 
\end{multline}  

\begin{multline}  
\bm{x}(k + i \mid k) \in E, \Delta \bm{u} \in \Delta U, \bm{u}(k + i \mid k) \in U, \\ i = 1, \ldots, N_p 
\end{multline}  

\begin{equation}  
\bm{x}_{a}^{p}(k + N_p \mid k) + \bm{x}_{a}^{p,\text{tilt}}(k + N_p \mid k) = \bm{x}_{a}^{a,\text{ref}}(k) 
\end{equation}  

\begin{equation}  
\bm{x}(k + N_p \mid k) = 0 
\end{equation}  
In the above equations, $N_p$ and $N_c$ denote the prediction and control horizon. $\bm{R}$, $\bm{S}$, and $\bm{Q}$ indicate the diagonal weighting matrices (DWMs) specific to the input rate, input, and output. $\bm{x}(k + i \mid k) \in E$, $\Delta \bm{u} \in \Delta U$ and $\bm{u}(k + i \mid k) \in U$ represent the limitation of the state vector, inputs' rates and inputs, respectively. $\bm{x}_{a}^{p} (k + N_p \mid k)$ and $\bm{x}_{a}^{p,\text{tilt}} (k + N_p \mid k)$ represent the predicted acceleration and tilt coordination acceleration of the vestibular system. $\bm{x}_{a}^{a,\text{ref}} (k)$ is the acceleration perceived by the aircraft pilot's vestibular system at the initial moment of the system. 

It is noted that $\bm{Q}_{N_p}$ is the terminal weighting matrix for the outputs using the terminal conditions and based on the Riccati equation, it can be calculated as follows:

\begin{multline}  
\bm{Q}_{N_p} = \bm{A}^T \bm{Q}_{N_p} \bm{A} - \bm{A} \bm{Q}_{N_p} \bm{B} \left( \bm{B}^T \bm{Q}_{N_p} \bm{B} + \bm{R} \right)^{-1} \\\bm{B}^T \bm{Q}_{N_p} \bm{A} + \bm{Q}
\end{multline}  

The acceleration experienced by the simulator pilot's vestibular system primarily arises from the acceleration produced by the simulator's linear movement and the acceleration induced by its tilting mechanism. Equation (25) ensures that the vestibular system of the simulator pilot experiences the same acceleration $\bm{x}_{a}^{a,\text{ref}}$, as that experienced by the aircraft pilot's vestibular system. To broaden the range of optimized control solutions, constraints are applied exclusively at the final stage of the cost function. The system's terminal state is denoted as $\bm{x}(k + N_p \mid k)$. An additional distinction between the MPC-based MCA without COTC and the MPC-based MCA with COTC lies in the endpoint state constraint, which must be zero, as defined in Equation (26).

\subsection{Controller design for  MPC based MCA without COTC}
The presence of terminal state constraints may render the optimisation problem infeasible. Consequently, it becomes imperative to transform optimisation problems that incorporate COTC into formulations devoid of these constraints to ensure solvability.

One difference between the MPC based MCA with COTC  and the MPC based MCA without COTC is that there are no state constraints at the terminal point. The other difference is that the TWMs of the existing MPC based MCA without COTC equal to the original weighting matrices $Q$ in equation (23) \cite{dagdelen2009model2}. The final optimisation problem is derived as:

\begin{multline}  
\min_{\substack{u(i), x(k) \in U}} J(k) = \sum_{i=1}^{N_p-1} \left( \bm{x}_{\text{ref}}(k + i \mid k) - \bm{x}(k + i \mid k) \right)^T \\\bm{Q} \left( \bm{x}_{\text{ref}}(k + i \mid k) - \bm{x}(k + i \mid k) \right)\\
+ \sum_{i=1}^{N_c-1} \bm{u}^T(k + i \mid k) \bm{S} \bm{u}(k + i \mid k) + \sum_{i=1}^{N_c-1} \Delta \bm{u}^T(k + i \mid k) \\\bm{R} \Delta \bm{u}(k + i \mid k) 
\end{multline}  
In the above equation, $\bm{R}$, $\bm{S}$, and $\bm{Q}$ denote the DWMs specific to the input rate, input, and output.

\subsection{Stability Analysis}

Cleij et al. (2019) and Rengifo (2018) \cite{cleij2019comparison,rengifo2018solving} developed a feedback control law tailored for linear systems, addressing the minimum energy regulator problem with fixed terminal constraints. This control law has been extended to linear time-varying systems using a receding horizon approach, which has been proven to ensure asymptotic stability. Our study employs this methodology to assess the stability of the MOC-based MCA. Additionally, the stability of the MCA without COTC can be guaranteed by utilizing a long prediction horizon, such as the infinite output prediction horizon \cite{rengifo2018solving}. The following equation illustrates the feedback control law:

\begin{equation}  
u(t) = -\bm{R}^{-1} \bm{B}^p \bm{P}^{-1}(t + T_s, t + N_p T_s) \bm{A} x(t) 
\end{equation}  

We adopted the Riccati equation for the calculation of  $\bm{P}^{-1}$:

\begin{multline}  
\bm{P}(t + k \mid t) = \bm{A}^{-1} \bm{P}(t + k + 1 \mid t) \bm{A}^{-1T} - \\\bm{A}^{-1} \bm{P}(t + k + 1 \mid t) \bm{A}^{-1T} \\
- \bm{A}^{-1T} \bm{C}^T \bm{E}^T \left[ \bm{I} + \bm{E} \bm{C} \bm{A}^{-1} \bm{P}(t + k + 1 \mid t) \bm{C}^T \bm{E}^T \right]^{-1} \\\bm{E} \bm{C} \bm{A}^{-1} \bm{P}(t + k + 1 \mid t) \bm{A}^{-1T} + \bm{B} \bm{R}^{-1} \bm{B}^T
\end{multline}  
Matrix $\bm{P}$ can be determined using Equation (29) and the inverse sum spanning from $N_p$ to a given time step. Additionally, $\bm{P}(t + T_s, t + N_p T_s)$ is set to zero as an additional constraint imposed on the system.

It is noted that the length of $N_p$ shall be sufficiently long enough to satisfy the requirement of the feedback control law in equation (30). It remains stable when the system in Equation (15) exhibits controllability and observability. Subsequently, when $A$ and $B$ are controllable, and $A$ and $C$ are observable, we adopt Theorem 1 and Theorem 2 from Kwon and Pearson \cite{kwon1978feedback} to validate whether the system is stable in Equation (15) under terminal states.

To verify the robustness of the proposed control system against lateral stall error injections, the following scenarios are taken into account which includes: violent turbulence, wind shear stall, lateral stall.

\subsection{Time Complexity Analysis}

The time complexity of the algorithm includes matrix multiplication and addition. These operations are usually \(O(n^2)\) to \(O(n^3)\) complexity, depending on the size of the matrix. However, since the size of these matrices is fixed (independent of \(N_p\)), we can consider that the time complexity of this part is fixed. 

In addition, the for-loop in the MPC algorithm, whose iteration times are controlled by \(N_p\) (variable), is used to build prediction models and optimize problems. The time complexity of this loop is \(O(N_p)\), because it contains a series of operations, and the number of these operations increases linearly with \(N_p\). 

Because the time complexity of matrix operations inside the loop is fixed, the time complexity of the whole algorithm is \(O(N_p)\).

\section{Simulation and Results}\label{section4}
\subsection{Simulation Setup}
Both the S-MPC based MCA with and without COTC are implemented in MATLAB to evaluate the controller performance. A virtual sensor is strategically placed at the pilot's eye point to record reference acceleration and angular velocity data. The recorded data is subsequently processed through a vestibular system model. The data structure consists of lateral and longitudinal acceleration information to be used as equality constraints, as well as angular velocity information and corresponding platform state values to resolve the optimisation problem. Comprehensive simulations and analyses were conducted to evaluate performance under distinct flight scenarios, specifically aeroplane turbulence and wing horizontal stall. It is found that the proposed approach outperforms MPC based MCA in terms of error rate by 25 percent (S-MPC based MCA  presents 0.0000125 error rate whilst MPC based MCA presents 0.00001 error rate).

The parameters used in flight simulator to demonstrate the algorithm performance are shown in Tab. \ref{table:performance}.

\begin{table}[htbp]
\centering
\begin{tabular}{|c|c|c|c|}
\hline
\textbf{} & \textbf{Excursion} & \textbf{Vel.} & \textbf{Acc.} \\
\hline
x & $\pm$1.7m & $\pm$1.5m/s & $\pm$10m/s2 \\
\hline
y & $\pm$1.7m & $\pm$1.5m/s & $\pm$10m/s2 \\
\hline
z & $[2.2,3.8]$m & $\pm$1m/s & $\pm$7m/s2 \\
\hline
row & $\pm$25$^\circ$ & $\pm$30$^\circ$/s & $\pm$200$^\circ$/s2 \\
\hline
pitch & $\pm$25$^\circ$ & $\pm$30$^\circ$/s & $\pm$200$^\circ$/s2 \\
\hline
yaw & $\pm$30$^\circ$ & $\pm$30$^\circ$/s & $\pm$200$^\circ$/s2 \\
\hline
$l_{1\ldots 6}$ & $[2.5,4.5]$m & & \\
\hline
\end{tabular}
\caption{6-DoF Steward simulator configurations.}
\label{table:performance}
\end{table}

\begin{figure*}[h]
\centering
\subfigure[ lateral acceleration  ]{
\begin{minipage}[t]{0.5\linewidth}
\centering 
\label{fig:8}
\includegraphics[width=3.5in,height=2in]{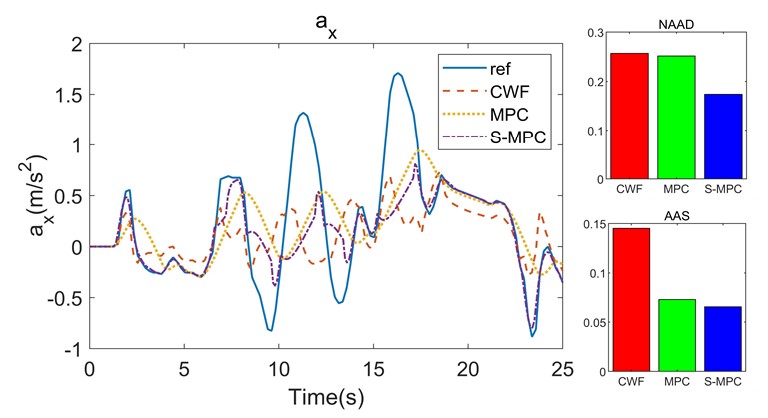}
\end{minipage}%
}%
\subfigure[longitudinal acceleration ]{
\begin{minipage}[t]{0.5\linewidth}
\centering 
\label{fig:9}
\includegraphics[width=3.5in,height=2in]{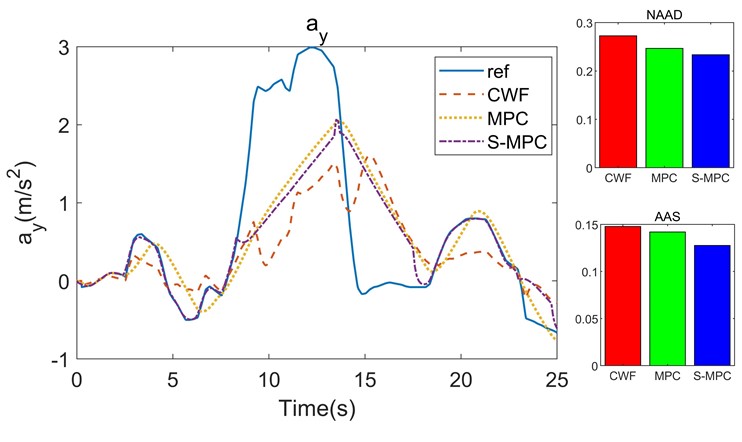}
\end{minipage}%
}%

\subfigure[ vertical acceleration ]{
\begin{minipage}[t]{0.5\linewidth}
\centering 
\label{fig:10}
\includegraphics[width=3.5in,height=2in]{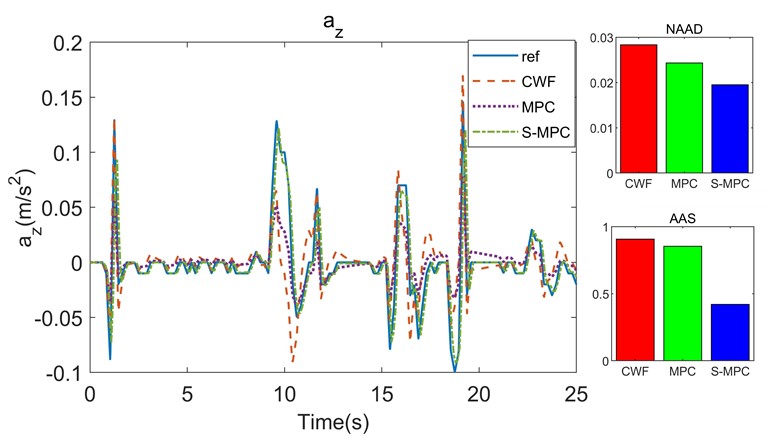}
\end{minipage}%
}%
\subfigure[roll rate ]{
\begin{minipage}[t]{0.5\linewidth}
\centering 
\label{fig:11}
\includegraphics[width=3.5in,height=2in]{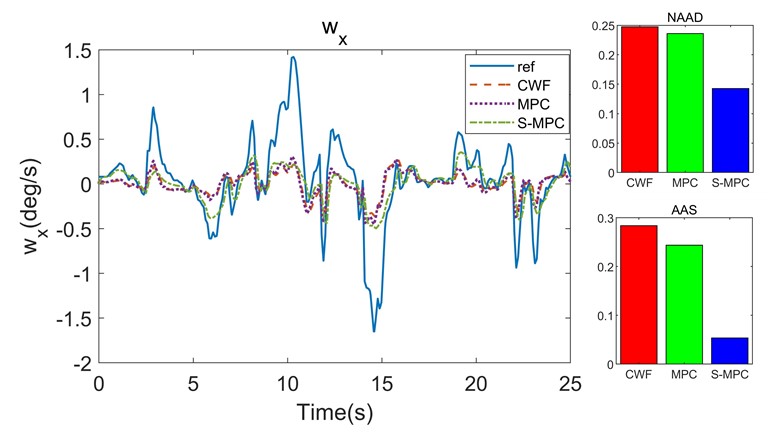}
\end{minipage}%
}%

\subfigure[pitch rate ]{
\begin{minipage}[t]{0.5\linewidth}
\centering 
\label{fig:12}
\includegraphics[width=3.5in,height=2in]{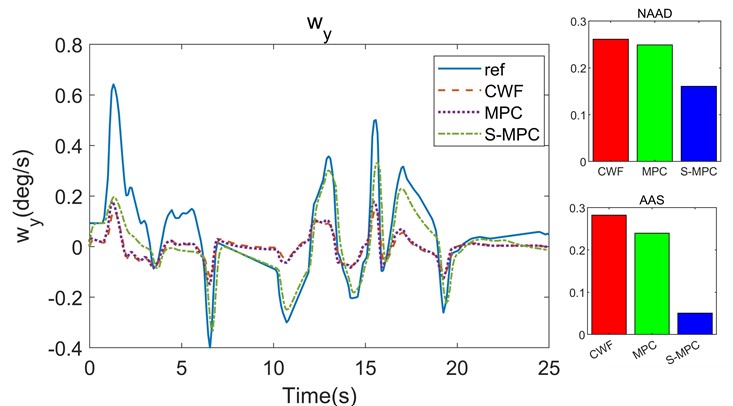}
\end{minipage}%
}%
\subfigure[yaw rate ]{
\begin{minipage}[t]{0.5\linewidth}
\centering 
\label{fig:13}
\includegraphics[width=3.5in,height=2in]{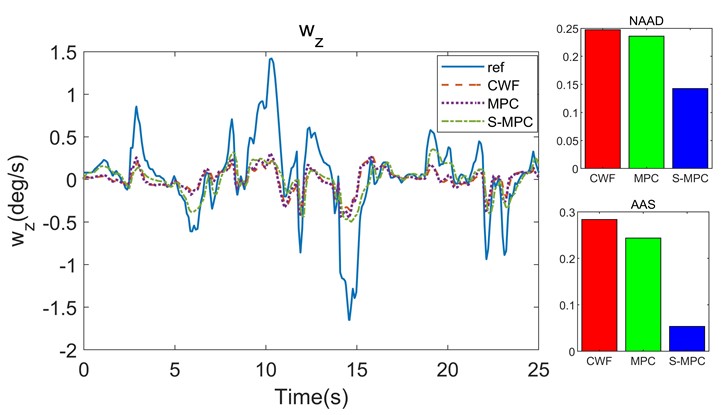}
\end{minipage}%
}%

\centering
\caption{Performance analysis of UPRT simulation results in the bumpy scenario.}
\end{figure*}

\subsection{S-MPC Performance over Bumpy Scenario}
Horizontal stall of aircraft wings is one typical failure of stall. When the wing stalls horizontally, the lateral acceleration of the aircraft will undergo drastic changes in a short period. The UPRT requires the simulator to replicate such drastic acceleration changes under the stall scenarios, requiring the somatosensory simulation algorithm to have fast response manoeuvrability and good somatosensory simulation performance. Fig. \ref{fig:8} demonstrates tracking performance towards lateral stall acceleration to verify the motion fidelity towards real conditions.

To compare performance, the Normalized Average Absolute Difference (NAAD) and Average Absolute Scale (AAS) metrics are used to evaluate the reference tracking capabilities of different MCAs. The performance comparison is carried out among S-MPC, MPC-based MCA, and CWF. Each algorithm has been carefully fine-tuned to optimize reference tracking accuracy while maintaining actuator positions within the defined physical limits.

From Fig. \ref{fig:8}, it is found that S-MPC based MCA has lower NAAD values compared to MPC based MCA and CWF. The NAAD and AAS values show that the S-MPC based MCA is superior to CWF and MPC, and during the time periods of 10s to 13s and 16s to 18s in the figure, S-MPC based MCA can switch controllers according to the physical limitations in the simulation platform, ensuring better tracking of targets by the simulation function.

Figure \ref{fig:9} demonstrates comparable findings, showing that the S-MPC-based MCA achieves superior tracking performance. This is evidenced by its relatively low NAAD and AAS values. In contrast, the tracking performance of CWF is once again found to be suboptimal when compared to the other two algorithms.

From Fig. \ref{fig:10}, S-MPC based MCA has better tracking capability versus the MPC based MCA and CWF. Still, its tracking accuracy is far inferior to SMPC based MCA.

From Figures \ref{fig:11}, \ref{fig:12}, and \ref{fig:13}, it is evident that due to the physical limitations of the simulation tracking platform, the tracking results for all three algorithms generally remain within the range of -0.5 deg/s to 0.5 deg/s. The S-MPC-based MCA demonstrates optimal approximation tracking within this range and outperforms the other algorithms. While CWF exhibits tracking accuracy comparable to that of the MPC-based MCA, it lacks the same level of adaptability and responsiveness evident in the tracking trend of the MPC-based MCA.

\subsection{S-MPC Performance over Horizontal Stall Scenario}
Horizontal stall of aircraft wings is the other common failure of stall, leading to drastic changes in the lateral acceleration of the aircraft. Fig. \ref{fig:8} demonstrates tracking performance towards horizontal stall acceleration to verify the motion fidelity towards real conditions.

\begin{figure}[htbp]
\centering
\caption{Performance analysis of lateral acceleration changes over different MCAs in UPRT under horizontal stall scenario.}
\includegraphics[scale=0.75]{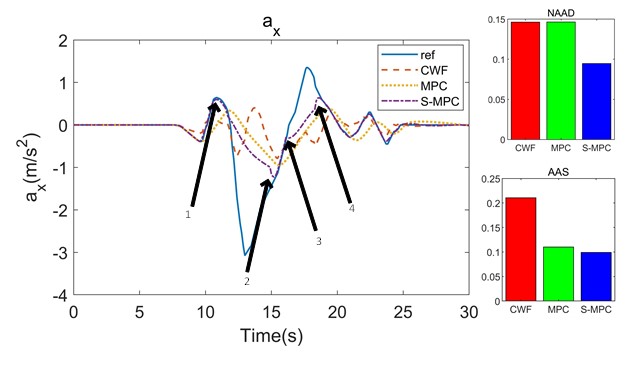}
\label{fig:14}
\end{figure}

\begin{figure}[htbp]
\centering
\caption{Performance analysis of roll rate changes over different MCAs in UPRT under horizontal stall scenario.}
\includegraphics[scale=0.75]{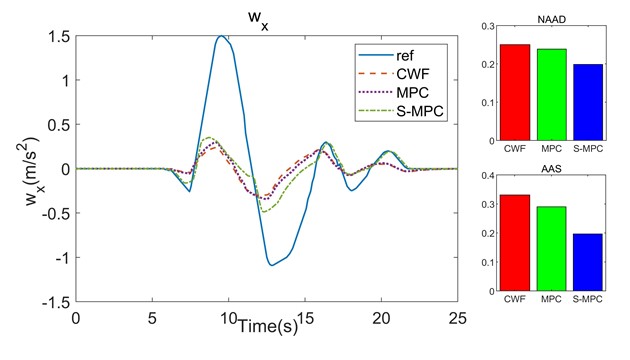}
\label{fig:15}
\end{figure}

As depicted in Fig. \ref{fig:14} and Fig. \ref{fig:15}, the S-MPC-based MCA exhibits a notable enhancement in tracking performance when compared to the CWF-based MCA and the MPC-based MCA, particularly in terms of its rapid convergence to reference trajectories. The transition between stages 1 and 2, as well as stages 3 and 4, is adeptly managed using MPC without COTC, ensuring optimal approximation tracking right at the edge of the simulator’s performance envelope. For the remainder of the simulation period, the adoption of MPC with COTC is crucial to achieving the best possible approximation tracking within the simulator’s operational envelope while also maintaining high fidelity in motion simulations. During the horizontal stall simulation, the AAS indices recorded for the S-MPC, MPC, and CWF algorithms are 0.196, 0.279, and 0.324, respectively, underscoring the superior performance of the S-MPC approach in maintaining stability and precision.

The S-MPC algorithm demonstrates significant improvements, outperforming the MPC and CWF algorithms by 42.34\% and 65.30\%, respectively.

\section{Conclusion}\label{section5}
A new S-MPC based MCA algorithm was proposed here, which integrated the vestibular system dynamic response and kinematic model towards the 6- DoF Stewart platform. The accurate and rapid tracking of high dynamic manoeuvres is achieved with MPC based MCA with COTC within the simulator operating envelope. The optimal tracking is enabled by switching between controllers, especially by using MPC based MCA without COTC when the operating envelope is identified to sit outside the simulator boundaries.

The proposed algorithm is evaluated under UPRT scenarios, and performance comparisons are conducted between the algorithm and the CWF and MPC based MCA. The MPC-based MCA outperforms CWF due to timely optimisation of the motion perception difference between the aircraft and the simulator. However, due to the MPC based MCA with scarce COTC constraints, the simulation platform capability is not maximised. The proposed method dynamically switches control algorithms to deploy the S-MPC-based MCA in real time, delivering exceptional simulation performance under large and overloaded UPRT conditions.




\end{document}